# Interfacial Charge Transfer and Morphology Govern Magnetism in Ultrathin $VO_x/MoSe_2$ Heterostructures

*Yasinthara Wadumesthri [1†], Nalaka Kapuruge [1†], Raphael B. de Oliveira [2,3], Bruno Ipaves [2], Guilherme S. L. Fabris [2], Kinga Lasek [1], Florence A. Nugera [1], Jianjun Pan [1], Xiaomei Jiang [1], Douglas S. Galvão [2*] and Humberto Rodríguez Gutiérrez [1*]*

[1] *Department of Physics, University of South Florida, Tampa, Florida 33620, USA*

[2] *Applied Physics Department and Center for Computational Engineering & Sciences, State University of Campinas, Campinas, Sao Paulo 13083-970, Brazil*

[3] *Department of Physics, Federal University of Rio Grande do Norte, Natal, Rio Grande do Norte 59078-970, Brazil.*

[†] *Equal contribution*

[*]*Corresponding Authors*: galvao@ifi.unicamp.br and humberto3@usf.edu

**Abstract**

Engineering magnetism in two-dimensional semiconductors through interfacial interactions offers an attractive alternative to substitutional doping and provides opportunities for integrating magnetic and optoelectronic functionality within the same heterostructure. Here, we investigate ultrathin $VO_x$ films deposited on monolayer $MoSe_2$ and demonstrate a pronounced enhancement of the magnetic response upon formation of the $VO_x/MoSe_2$ interface. Combined structural, spectroscopic, magnetic, and first-principles analyses support a consistent picture in which finite $VO_x$ clusters possess intrinsic magnetic moments that are further enhanced by electron transfer from $MoSe_2$ to the oxide. The predicted charge transfer is also consistent with the increased trion contribution observed in the photoluminescence response of the heterostructure. Beyond this interfacial enhancement, the magnetic behavior is strongly governed by the morphology of the ultrathin oxide. A crossover from discontinuous clusters to more continuous films produces qualitatively different temperature dependences: the continuous-film regime exhibits conventional thermal demagnetization and coercive softening, whereas the clustered regime displays an unusual increase in magnetization and nonmonotonic coercivity with increasing temperature. These contrasting responses are captured by a phenomenological model in which morphology determines the collective magnetic dynamics through a distribution of effective reversal barriers and their thermally activated accessibility. Together, the results establish a unified picture in which interfacial charge transfer enhances the local magnetic moments of $VO_x$, while nanoscale morphology governs their collective magnetic dynamics, providing complementary routes for controlling magnetism in oxide/transition-metal dichalcogenide heterostructures.

The ability to engineer magnetic functionality in atomically thin materials has emerged as a central objective in condensed matter physics and nanotechnology because of its potential impact on spintronic, magneto-optical, and quantum information devices.[1-3] Two-dimensional transition-metal dichalcogenides (TMDs) are particularly attractive owing to their strong spin-orbit coupling, direct band gaps in the monolayer limit, robust excitonic effects, and compatibility with van der Waals heterostructures.[4-7] Considerable effort has therefore focused on introducing magnetic functionality into otherwise nonmagnetic TMDs while preserving their structural and optical quality.[8, 9]

Several strategies have been explored to realize magnetism in TMDs, including substitutional magnetic doping,[10-17] defect engineering,[18, 19] proximity coupling to magnetic materials,[20-22] and intrinsically magnetic van der Waals crystals.[2, 3, 23-25] Although these approaches have produced important advances, they can introduce limitations such as chemical disorder, degraded carrier transport, limited ordering temperatures, or restricted material choices. These challenges motivate alternative routes for modifying magnetic properties without substantially disrupting the host semiconductor lattice.[26]

Interface engineering has recently emerged as one such alternative.[27-29] Electronic charge redistribution across oxide/semiconductor interfaces can modify orbital occupation, oxidation state, and exchange interactions without requiring extensive atomic interdiffusion.[30] Transition-metal oxides are particularly attractive because their multiple oxidation states and correlated electronic structure make their magnetic properties highly sensitive to their local electronic environment.[31-33] Among transition-metal oxides, vanadium oxides are especially versatile because vanadium can adopt multiple oxidation states, while the V–O system supports numerous stoichiometric, oxygen-deficient, mixed-valence, and metastable phases with diverse crystal structures and electronic properties.[34] While bulk $V_2O_5$ is generally nonmagnetic, reduced coordination, oxygen vacancies, finite clusters, and non-stoichiometric $VO_x$ nanostructures can develop localized magnetic moments.[35-37] Recent first-principles calculations on a $V_2O_3/WSe_2$ van der Waals heterostructure further illustrate this potential, predicting strong interfacial charge redistribution and magnetic proximity coupling between a two-dimensional vanadium oxide and a TMD semiconductor.[38] Consequently, ultrathin $VO_x$ deposited on a two-dimensional semiconductor provides an excellent platform for investigating how interfacial electronic interactions modify pre-existing magnetic moments while preserving the optical and electronic properties of the underlying 2D semiconductor.

Motivated by our preliminary experimental results,[39] Woods' group employed density-functional-theory calculations to investigate an idealized single-crystal $V_2O_5/MoSe_2$ interface, predicting that interfacial charge transfer, orbital hybridization, and spin redistribution can promote ferromagnetism.[40] Together with recent theoretical predictions for $V_2O_3/WSe_2$ heterostructures,[38] these studies highlight the potential of vanadium-oxide/TMD interfaces for engineering magnetic functionality. However, the relationship among charge transfer, optical response, morphology, and magnetic behavior in experimentally realized nonstoichiometric $VO_x$/TMD heterostructures remains largely unexplored. Moreover, the role of oxide morphology—particularly the transition from isolated nanoclusters to continuous ultrathin films—in determining collective magnetic behavior remains poorly understood.[41]

In this work, we investigate ultrathin $VO_x/MoSe_2$ heterostructures in which the oxide morphology evolves from discontinuous islands to laterally connected ultrathin films with increasing thickness. By combining atomic force microscopy, Raman and photoluminescence spectroscopy, X-ray photoelectron spectroscopy, vibrating-sample magnetometry, first-principles calculations, and a phenomenological description of the temperature-dependent magnetic response, we establish a coherent physical picture. Our results indicate that finite $VO_x$ clusters possess intrinsic localized magnetic moments that are enhanced through interfacial electron transfer from the $MoSe_2$ substrate. Independently, the oxide morphology governs the collective magnetic dynamics by modifying the distribution of effective reversal barriers and magnetic interactions, producing qualitatively different temperature dependences for clustered and continuous films. This unified framework links the structural, optical, electronic, and magnetic properties of the heterostructure and illustrates how interface engineering can be exploited to tailor magnetic functionality in oxide/semiconductor van der Waals systems.

## Results and Discussion

Monolayer $MoSe_2$ samples were synthesized on Si substrates, covered with a 285 nm $SiO_2$ layer, using the water-assisted chemical vapor deposition (CVD) approach developed by our group.[42-44] The process is schematically illustrated in Figure 1a; additional synthesis details are provided in the Experimental Section. After CVD growth of $MoSe_2$, the samples were transferred to a high-vacuum thermal evaporator, where ultrathin $VO_x$ films were deposited on the $MoSe_2$ surface (Figure 1b). The nominal $VO_x$ thickness was monitored using an in-situ quartz crystal microbalance (QCM).

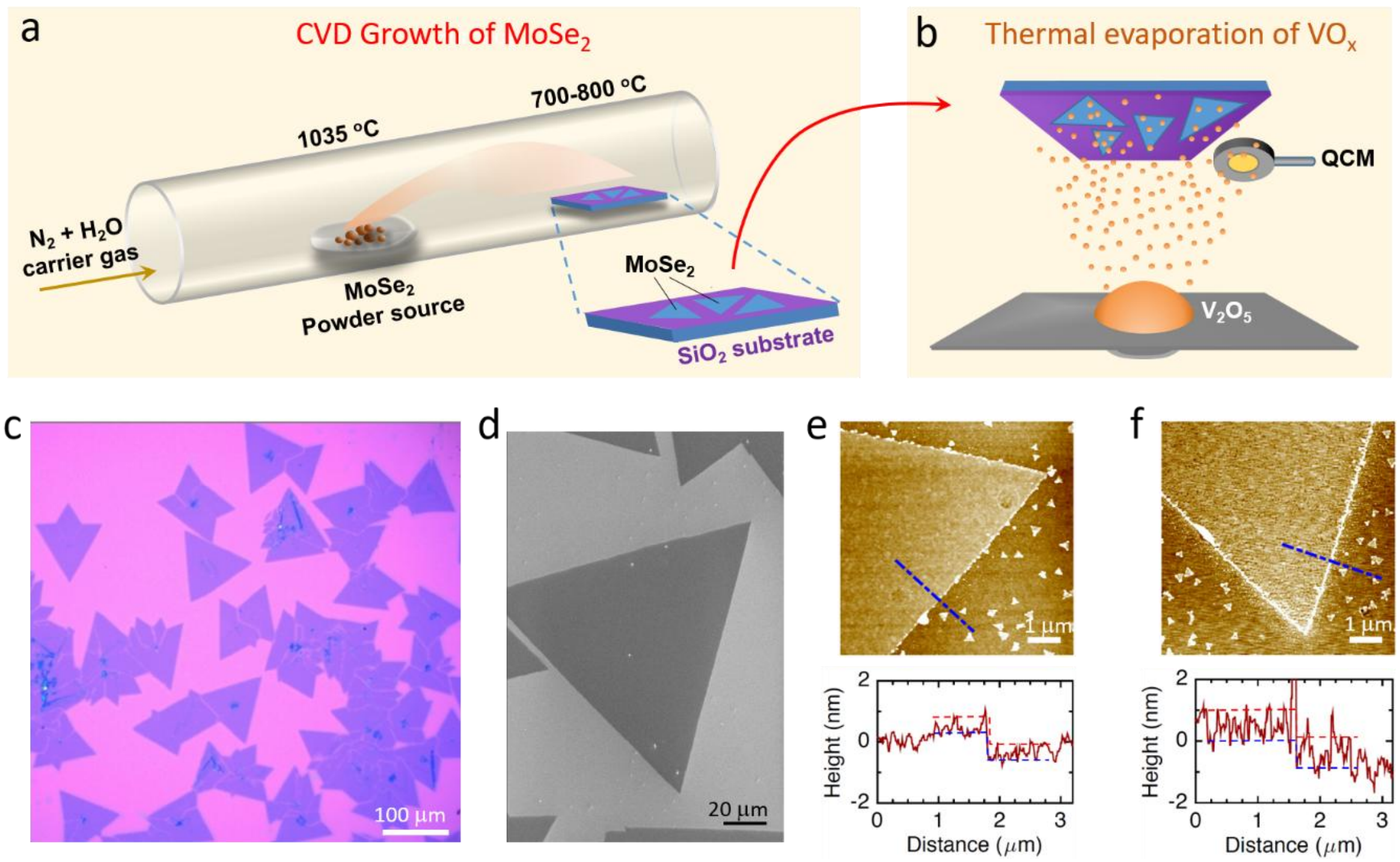

**Figure 1.** (a) Chemical vapor deposition of monolayer $MoSe_2$. (b) Thermal deposition of ultrathin $VO_x$ films. (c, d) Optical and SEM images, respectively, of the CVD-grown $MoSe_2$. (e, f) Atomic force microscopy (AFM) images (top) and height profiles (bottom) of $MoSe_2$ before and after deposition of a nominal 0.3 nm $VO_x$ layer, respectively.

Optical microscopy (Figure 1c) and scanning electron microscopy (SEM) (Figure 1d) reveal triangular CVD-grown $MoSe_2$ islands with a relatively high surface density. An atomic force microscopy (AFM) image and height profile of pristine $MoSe_2$ are shown in Figure 1e. The measured thickness is approximately 0.7 nm, consistent with monolayer $MoSe_2$, and the average roughness is below 0.45 nm, comparable to that of the $SiO_2$ substrate. After deposition of a nominal ~0.3 nm $VO_x$ layer, as determined by the QCM, the average roughness approximately doubles (Figure 1f). The resulting room-temperature-deposited oxide is therefore consistent with a discontinuous nanoparticle-like morphology.

Using a shadow mask, step-like $VO_x$ films with different nominal thicknesses were deposited directly on $SiO_2$ substrates (Figure 2). *Ex situ* AFM step-height measurements performed in tapping mode under ambient conditions yielded values slightly larger than the nominal QCM thicknesses (Figure 2e). The near-unity slope and small 0.3 nm intercept suggest a systematic offset that may arise from material-dependent apparent-height effects, because tapping-mode AFM step heights across dissimilar materials can depend on tip–sample interactions, capillary forces, adsorbed water layers, and measurement parameters.[45-48] Accordingly, QCM values are used throughout this work as the nominal film thickness. The AFM analysis further indicates a thickness-dependent morphological evolution of the vanadium oxide layer. At the lowest nominal thickness, the greater roughness of the $VO_x/SiO_2$ region, relative to bare $SiO_2$, is consistent with an island-like or discontinuous oxide morphology. With increasing nominal thickness, the roughness decreases towards that of the substrate, suggesting a transition to a more continuous film. This morphological evolution is relevant to the magnetic response discussed below: discontinuous $VO_x$ clusters in the ultrathin regime exhibit finite-size and thermally activated behavior, whereas thicker, more continuous films are expected to approach the response of an extended oxide layer.

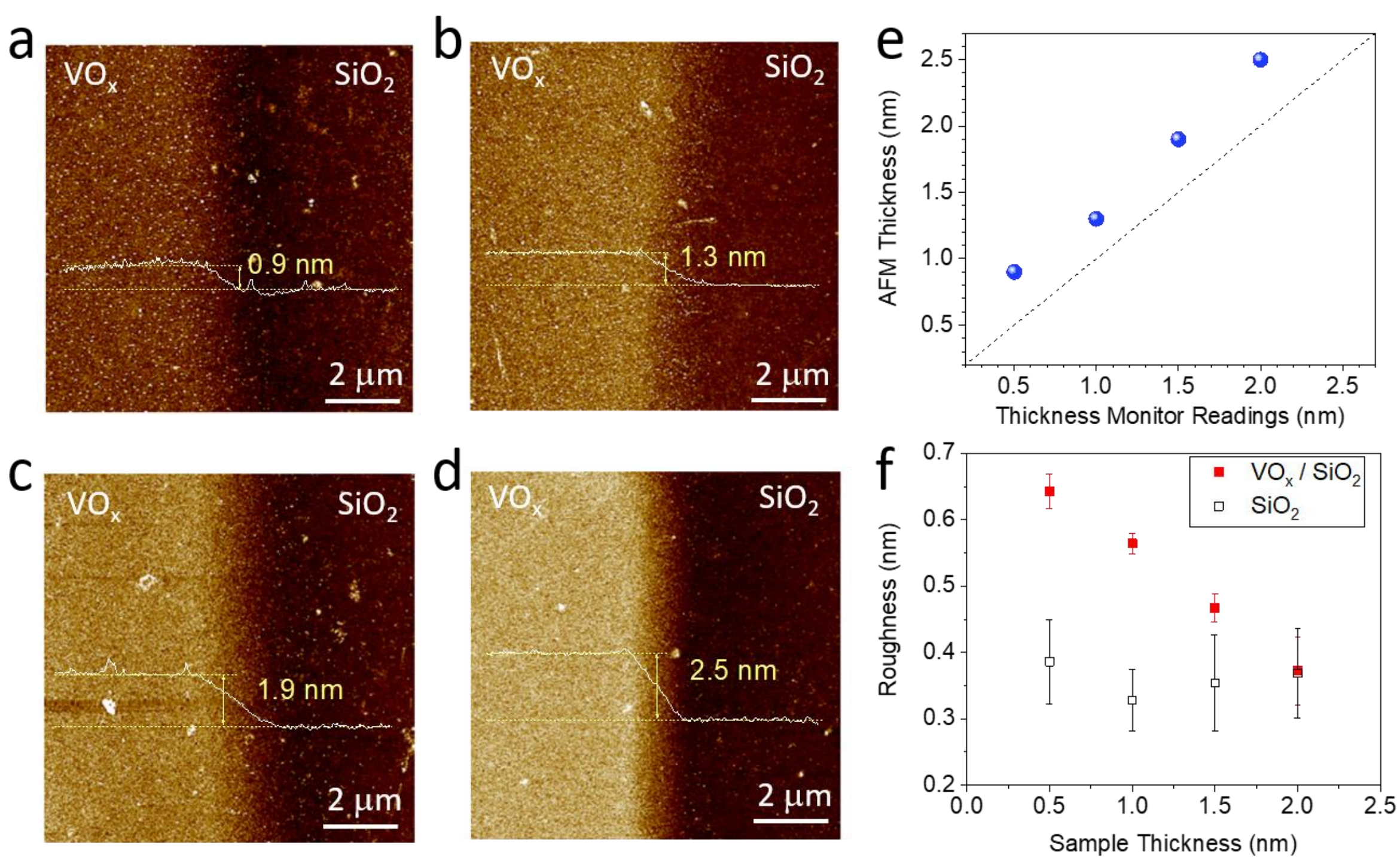

**Figure 2.** (a - d) AFM topography images and corresponding averaged step profiles for $VO_x$ films deposited on $SiO_2$ substrates. (e) Correlation between film thickness measured by AFM and the nominal thickness measured by the in-situ QCM. (f) Surface roughness as a function of nominal film thickness determined from the AFM images in panels a-d.

Next, we compare the optical properties of pristine $MoSe_2$ with those of a heterostructure containing a nominal 0.7 nm $VO_x$ overlayer. The Raman spectra are shown in Figure 3a. Two prominent peaks near 239 and 249 $cm^{-1}$ are assigned to the $A_{1g}$ and $2E_{2g}^2(M)$, respectively.[49] Additional weaker features at 146, 170, 286, 305, 319 and 361 $cm^{-1}$ are assigned to $E_{2g}^1$-LA(M), $E_{1g}$, $E_{2g}^1$, $E_{1g}$+LA(M), $2B_{2g}(M)$ and $A_{2u}^2$, respectively, following Nam et al.[49] The close similarity in peak positions and relative intensities before and after $VO_x$ deposition indicates that the overlayer produces no readily detectable change in the crystalline structure of the underlying $MoSe_2$.

The photoluminescence (PL) spectrum of pristine $MoSe_2$ exhibits an intense peak at 1.53 eV (Figure 3b,c), consistent with our previous reports.[42-44] The spectrum is slightly asymmetric toward lower energy and can be fitted with two Gaussian components (Figure 3c): a dominant neutral-exciton peak at 1.53 eV and a weaker trion contribution near 1.48 eV. After $VO_x$ deposition, the PL maximum red-shifts slightly and the low-energy side broadens. For the nominal 0.7 nm $VO_x/MoSe_2$ sample, the red shift is less than 15 meV. Such a shift could arise from tensile strain[50, 51] or a modified dielectric environment.[52] Because the Raman spectra show negligible strain-related changes, the PL red shift is more plausibly associated with the altered dielectric environment produced by the $VO_x$ overlayer.

The nominal 0.7 nm $VO_x/MoSe_2$ sample likewise exhibits an asymmetric PL peak that can be fitted with two Gaussian components (Figure 3d), centered at 1.515 and 1.475 eV and assigned to neutral excitons and trions, respectively. In both pristine and coated $MoSe_2$, the neutral-exciton–trion energy separation is 40-50 meV, close to the room-temperature value reported previously.[53, 54] In addition, the trion-to-neutral-exciton intensity ratio ($I_t/I_x$) increases from 0.14 in pristine $MoSe_2$ to 0.37 after $VO_x$ deposition. This increase could arise from charge transfer between the $VO_x$ overlayer and $MoSe_2$ or from V incorporation into the $MoSe_2$ lattice; substitutional V at Mo sites is expected to induce p-type doping.[10] Either mechanism would alter the free-carrier concentration and, consequently, the trion population. The XPS results below allow these possibilities to be distinguished.

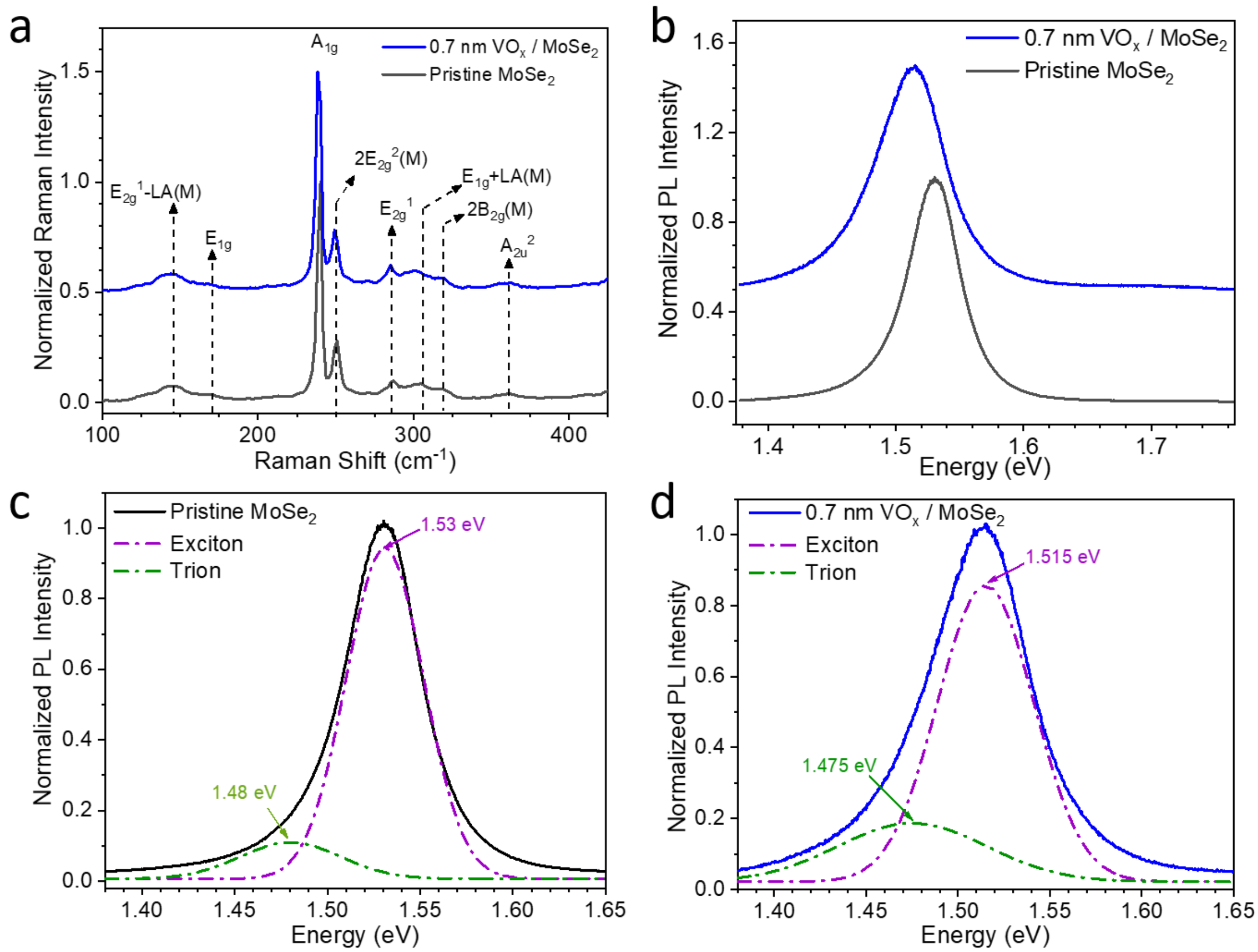

**Figure 3**. (a, b) Raman and PL spectra, respectively, of pristine $MoSe_2$ and nominal 0.7 nm $VO_x/MoSe_2$ heterostructure. (c, d) PL spectra of pristine $MoSe_2$ and 0.7 nm- $VO_x/MoSe_2$, respectively, showing the fitted neutral-exciton and trion contributions.

X-ray photoelectron spectroscopy (XPS) was used to assess whether V deposition produces measurable interdiffusion or substitutional incorporation into the $MoSe_2$ layer. As shown in Figure 4a, the Mo 3d and Se 3d spectra of pristine $MoSe_2$ and $VO_x$ /$MoSe_2$ are nearly identical, with no significant core-level shifts or obvious changes in line shape. Previous work reported that even ~1% substitutional V doping in $MoSe_2$ can induce binding-energy shifts of approximately 0.5 eV in the Mo 3d and Se 3d levels.[10] The absence of comparable shifts in our samples indicates that substitutional V incorporation is negligible or below the detection limit of the XPS measurements. In contrast, the V 2p spectrum of the oxide layer contains contributions from $V^{5+}$ and reduced V species, indicating a deviation from ideal $V_2O_5$ stoichiometry. These results support a heterostructure consisting of $MoSe_2$ interfaced with a mixed-valence, nonstoichiometric $VO_x$ overlayer, rather than a significantly V-doped $MoSe_2$ layer.

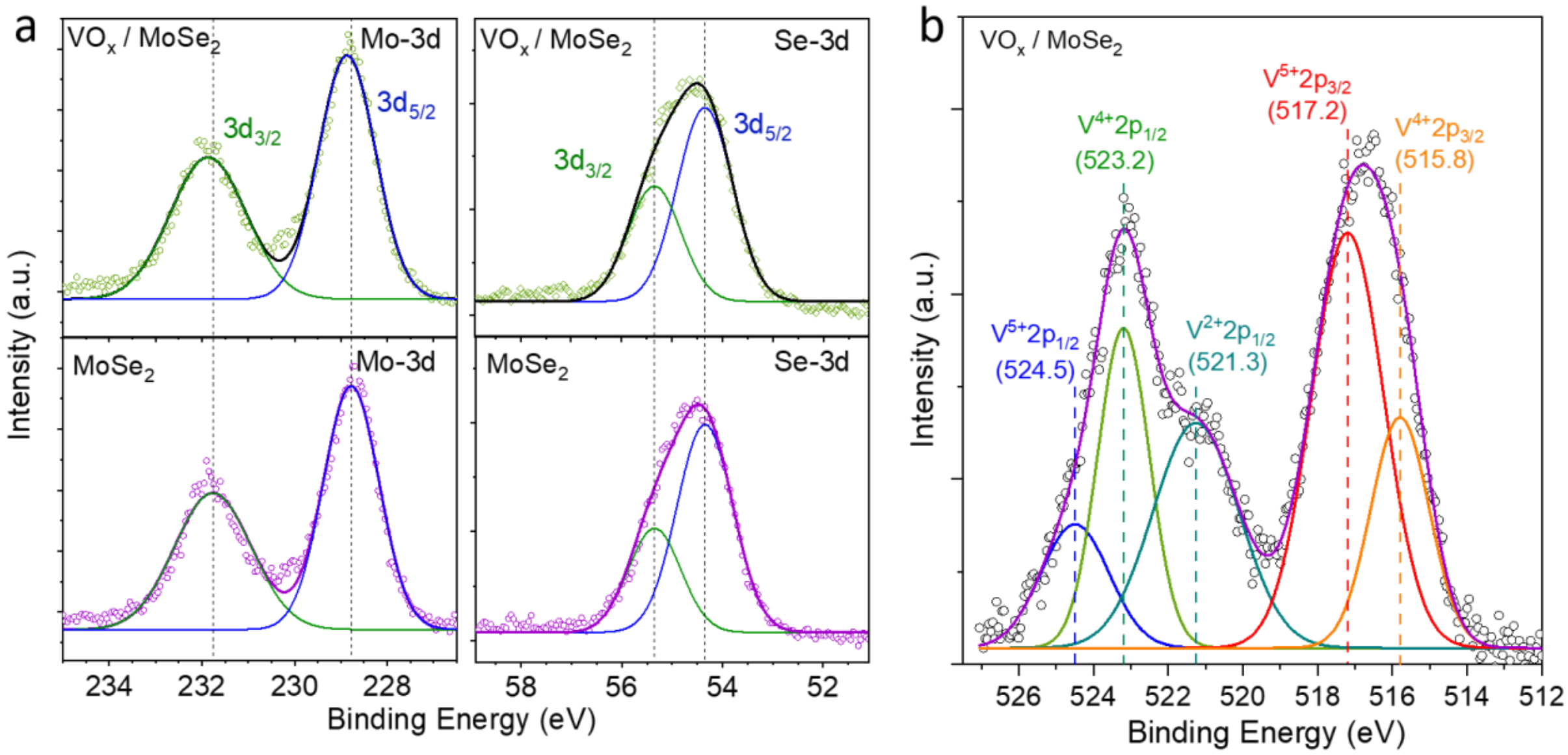

**Figure 4**. XPS spectra of (a) the Mo 3d and Se 3d regions for pristine $MoSe_2$ and $VO_x/MoSe_2$ and (b) the V 2p region for $VO_x/MoSe_2$.

The magnetic properties of the $VO_x/MoSe_2$ heterostructures were investigated using the vibrating sample magnetometer (VSM) option of a Quantum Design Physical Property Measurement System (PPMS). Figure 5 summarizes the areal magnetization as a function of magnetic field (M-H) after subtraction of the diamagnetic background. The top and bottom panels correspond to magnetic fields applied in-plane and out-of-plane, respectively. Pristine $MoSe_2/SiO_2$ and $VO_x/SiO_2$ show no detectable hysteretic response for an in-plane field (Figure 5a), whereas a weak out-of-plane response appears for the $VO_x/SiO_2$ films thicker than 1 nm (Figure 5d). In contrast, all $VO_x/MoSe_2$ heterostructures exhibit clear hysteresis, with a substantially stronger response in the out-of-plane configuration.

The enhanced ferromagnetic response observed after depositing vanadium oxide on $MoSe_2$ suggests that the interface actively modifies the magnetic properties of the oxide layer. To gain microscopic insight into this enhancement, we performed first-principles DFT+U calculations on representative $VO_x$ nanoclusters ($V_4O_{10}$ and $V_{10}O_{22}$) interacting with a $MoSe_2$ monolayer; computational details and additional results are provided in the Supporting Information. The calculations represent the early stages of film growth, in which isolated oxide clusters dominate before coalescence into a more continuous film, consistent with the AFM observations. Although only representative cluster geometries were considered, they capture local electronic interactions at the $VO_x/MoSe_2$ interface and provide a microscopic framework for interpreting the experimentally observed magnetic enhancement.

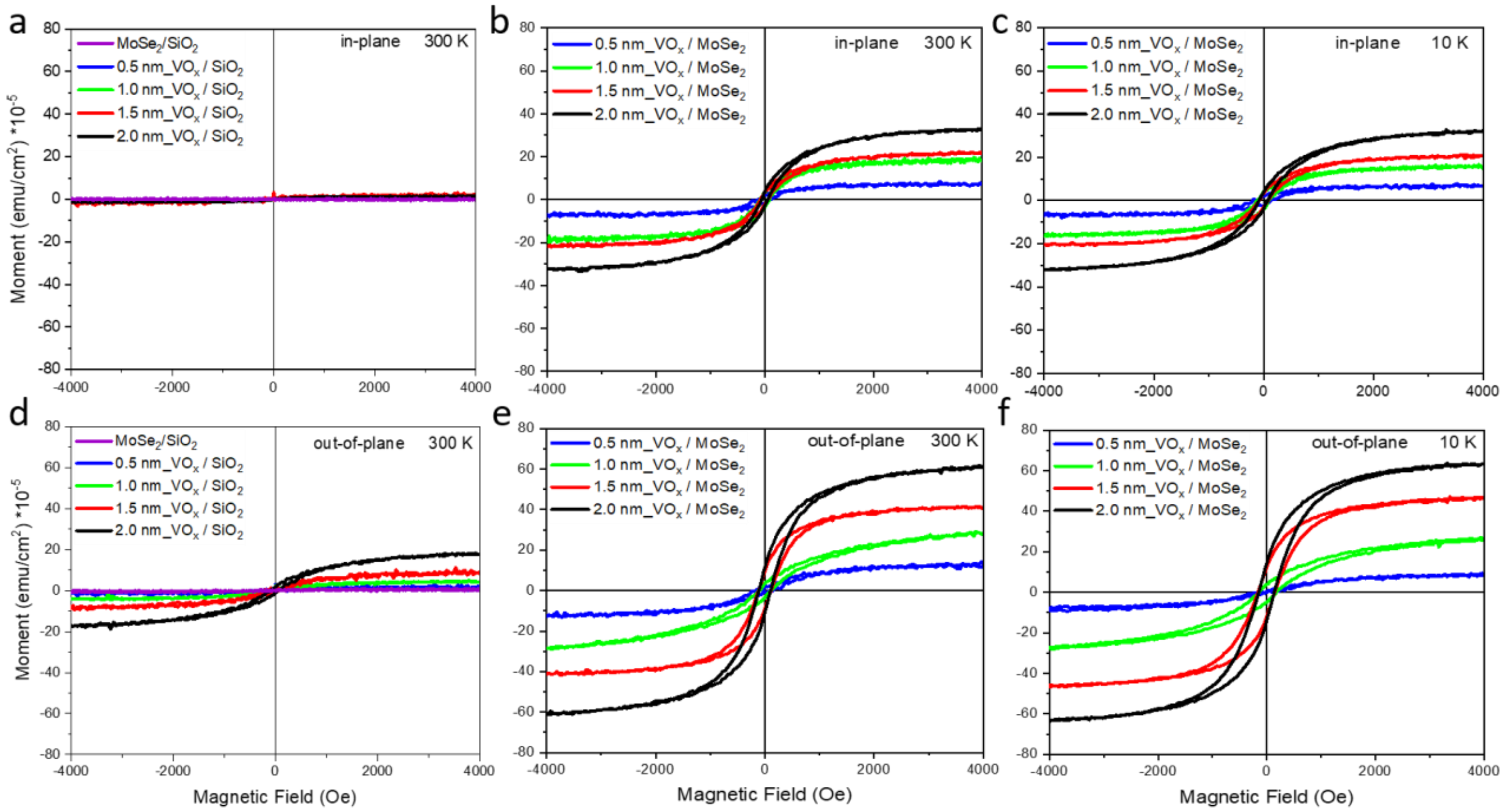

**Figure 5.** Magnetic properties of the individual constituents and $VO_x/MoSe_2$ heterostructures, with the external magnetic field applied in-plane (top panels) and out-of-plane (bottom panels). (a,d) Magnetic response of pristine $MoSe_2/SiO_2$ and $VO_x/SiO_2$ films with different nominal $VO_x$ thicknesses. (b, e) Hysteresis loops measured at 300 K and (c,f) at 10 K for $VO_x/MoSe_2/SiO_2$ heterostructures with different nominal $VO_x$ thicknesses.

Both pristine $MoSe_2$ and bulk $V_2O_5$ exhibit negligible calculated magnetic moments, consistent with the nonmagnetic character expected for these reference systems.[40] In contrast, finite $VO_x$ clusters develop localized magnetic moments upon structural relaxation. As illustrated for $V_4O_{10}$ in Figure 6b,c, relaxation modifies the local coordination of the finite oxide cluster and produces incomplete compensation between positive and negative spin contributions, resulting in a net magnetic moment of approximately $1.20\mu_B$. The corresponding spin-density distribution (Figure 6d) shows that this magnetic polarization is localized within the finite oxide cluster. A substantially larger moment of approximately $7.0\mu_B$ is obtained for the isolated $V_{10}O_{22}$ cluster, further demonstrating that finite vanadium oxide nanostructures can possess intrinsic magnetic moments even though bulk $V_2O_5$ is nonmagnetic.

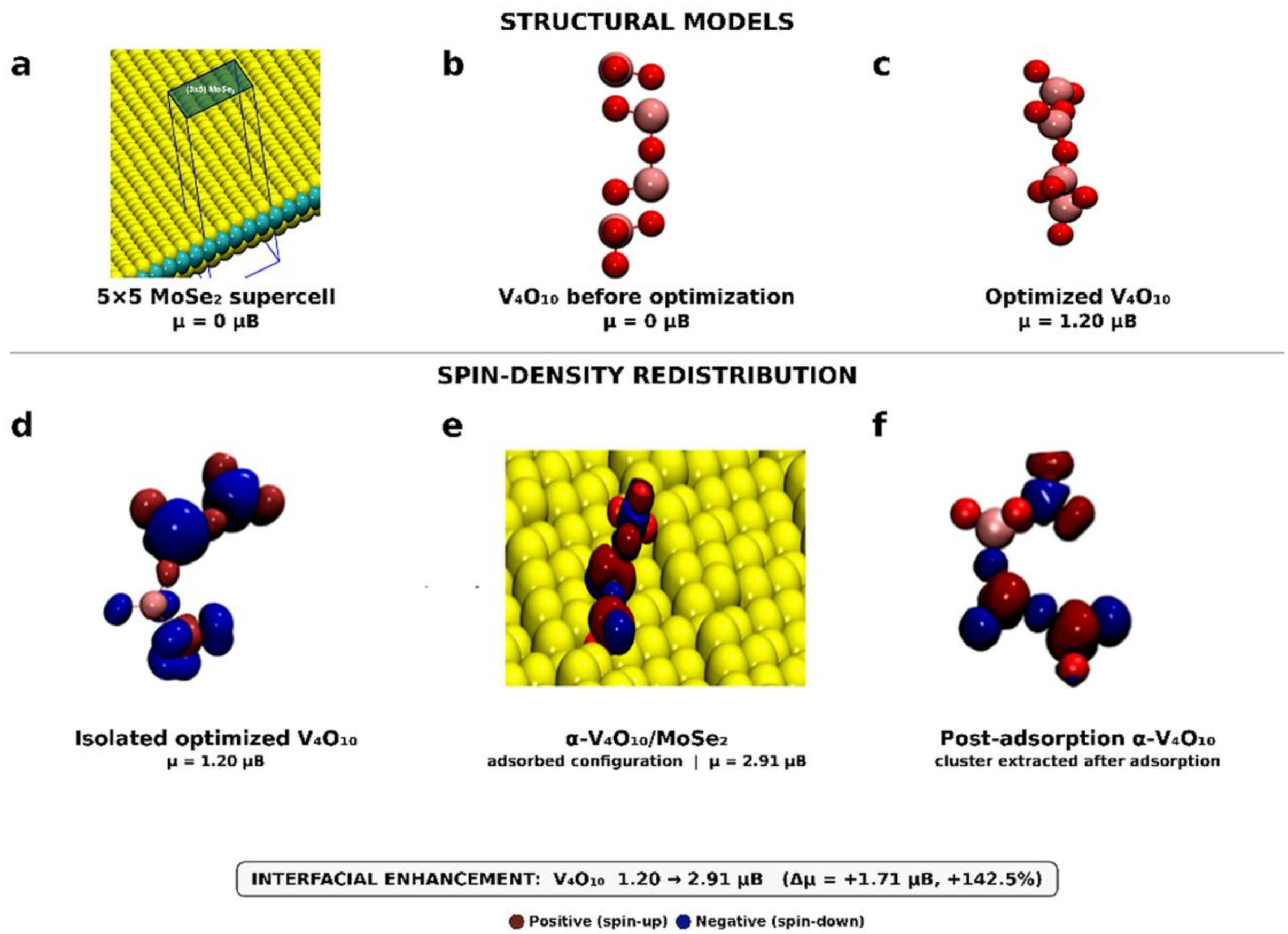

**Figure 6**. Structural evolution and spin-density redistribution in the $V_4O_{10}/MoSe_2$ system. (a) 5×5 $MoSe_2$ supercell employed to construct the hybrid structures. (b) Initial bulk-derived $V_4O_{10}$ cluster prior to structural relaxation and (c) optimized isolated $V_4O_{10}$ cluster. Structural relaxation results in incomplete compensation between local spin contributions and produces a finite magnetic moment of $1.20\mu_B$. (d) Spin-density distribution of the optimized isolated $V_4O_{10}$ cluster. (e) Spin-density distribution of the $\alpha$-$V_4O_{10}/MoSe_2$ configuration after adsorption, for which the total magnetic moment increases to $2.91\mu_B$. (f) Spin-density distribution of the $V_4O_{10}$ cluster extracted from the optimized heterostructure while retaining its post-adsorption geometry. Brown and blue isosurfaces represent positive (spin-up) and negative (spin-down) spin densities, respectively.

Interaction with $MoSe_2$ further modifies this spin-polarized state. As shown in Figure 6e, adsorption of the α-$V_4O_{10}$ cluster produces a pronounced redistribution of spin density within the oxide, concomitant with an increase in the total magnetic moment from 1.20 to 2.91$\mu_B$, corresponding to an enhancement of approximately 142.5%. Charge-density analysis further shows substantial electronic redistribution in the interfacial region, involving Se–O bonding, consistent with charge transfer between $MoSe_2$ and the oxide. Importantly, the enhanced spin polarization remains predominantly localized within $V_4O_{10}$ rather than being transferred to the $MoSe_2$ layer itself.

To further distinguish the effect of the relaxed adsorption geometry from the direct electronic contribution of the substrate, the $V_4O_{10}$ cluster was subsequently extracted from the optimized heterostructure while retaining its post-adsorption geometry (Figure 6f). Comparison of Figures 6d–f therefore allows the spin redistribution associated with structural relaxation and that arising directly from interfacial electronic coupling to be assessed separately. A similar interfacial enhancement is obtained for the larger $V_{10}O_{22}$ cluster, whose calculated magnetic moment increases from approximately 7.0 to 8.0$\mu_B$ after adsorption. Although the relative enhancement is smaller than for $V_4O_{10}$, this result indicates that the interfacial modulation of the oxide magnetic state is not restricted to a single cluster size or geometry.

Taken together, these calculations indicate that $MoSe_2$ does not become substantially magnetic upon contact with $VO_x$; instead, interfacial electronic reconstruction enhances magnetic moments that already emerge in finite oxide clusters, providing a microscopic explanation for the increased magnetic response observed experimentally in the $VO_x/MoSe_2$ heterostructures.

The calculated direction of charge transfer also provides a consistent interpretation of the optical measurements. Electron transfer from $MoSe_2$ to $VO_x$ corresponds to hole doping of the semiconductor and is therefore consistent with an increased contribution from positively charged trions. The PL measurements alone do not distinguish positive from negative trions, so this assignment is made only in conjunction with the independently calculated direction of charge transfer.

The thickness-dependent temperature evolution of magnetic properties was also investigated by VSM. Figure 7 summarizes the out-of-plane saturation magnetization ($M_s$) and coercive field ($H_c$). Panels 7a and 7b show $M_s$ and $H_c$, respectively, as functions of nominal $VO_x$ thickness between 10 and 350 K, whereas panels 7c and 7d show their temperature dependences for representative 1.5 and 0.5 nm samples.

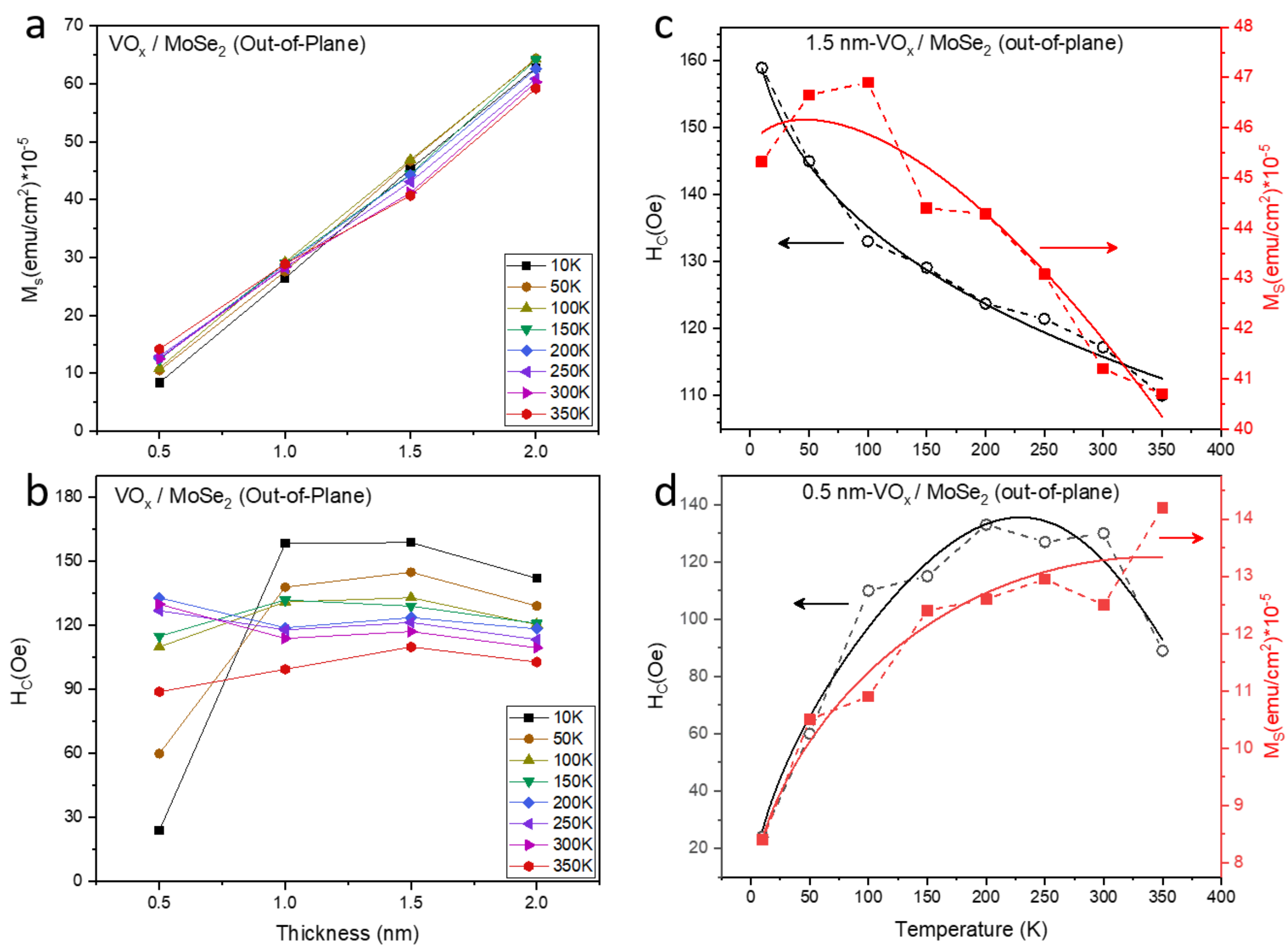

**Figure 7.** (a,b) Saturation magnetization ($M_s$) and coercive field ($H_c$), respectively, as functions of nominal $VO_x$ thickness at different temperatures. (c,d) $M_s$ (squares, right axes) and $H_c$ (circles, left axes) as functions of temperature for nominal $VO_x$ thicknesses of 1.5 and 0.5 nm, respectively. Solid lines are fits using the phenomenological model described in the text.

A pronounced change in the magnetic behavior is observed as the $VO_x$ thickness decreases below approximately 1 nm. The 1.5 nm sample exhibits the conventional temperature dependence expected for a continuous magnetic film. The saturation magnetization decreases gradually from approximately $47\times10^{-5}$ to $41\times10^{-5}$ emu·cm$^{-2}$ as the temperature increases from 10 to 350 K, while the coercive field decreases nearly monotonically from approximately 160 to 110 Oe. Such behavior is consistent with the progressive thermal excitation of spin fluctuations together with thermal activation over magnetic anisotropy barriers.

In striking contrast, the ultrathin 0.5 nm sample exhibits the opposite trend. The saturation magnetization increases continuously with increasing temperature, rising from $8.4\times10^{-5}$ to $14.2\times10^{-5}$ $emu\cdot cm^{-2}$, while the coercive field increases rapidly between 10 K and approximately 200 K before remaining nearly constant and eventually decreasing at the highest measured temperature. The coexistence of increasing magnetization and increasing coercivity with temperature is highly unusual for conventional ferromagnets and therefore suggests that different collective magnetic dynamics dominate in the ultrathin limit.

These results indicate that the temperature-dependent magnetic response changes qualitatively with oxide morphology. The AFM roughness measurements (Figure 2f) are consistent with discontinuous islands at subnanometer $VO_x$ coverage and a more laterally connected morphology at larger thickness. Accordingly, the contrasting $M_s(T)$ and $H_c(T)$ responses of the 0.5 and 1.5 nm samples cannot be described by ordinary thermal demagnetization alone; they also reflect morphology-dependent reversal dynamics.

We therefore use a phenomenological model to describe the temperature-dependent magnetic properties in Figure 7c,d. The model treats the charge-transfer-enhanced moments identified by DFT as pre-existing magnetic entities once the heterostructure is formed and assigns the contrasting temperature trends to differences in their collective dynamics. In this description, interfacial electronic reconstruction sets the magnetic-moment scale, whereas morphology controls the distribution of anisotropy- and interaction-modified reversal barriers.

This interpretation follows the established framework for fine magnetic particles and nanoparticle assemblies. In the Néel-Brown picture, a single-domain magnetic moment reverses by thermally activated crossing of an anisotropy barrier, with a characteristic relaxation time $\tau = \tau_0 e^{(E/k_B T)}$.[55, 56] Consequently, whether a moment appears blocked or dynamically responsive depends on the relation between its relaxation time and the experimental observation time. In a real cluster ensemble, variations in cluster size, shape, local anisotropy, defect structure, and orientation produce a distribution of barriers rather than one unique barrier.[57-59] Interparticle dipolar or exchange interactions further reshape this energy landscape and can increase, decrease, or broaden the effective barriers governing relaxation and hysteresis.[58, 60-63] Experimental studies of nanoparticle assemblies have directly shown that interactions and size dispersion modify blocking behavior, relaxation, and coercivity.[60, 61, 64, 65]

For the discontinuous subnanometer $VO_x$ film, the AFM morphology naturally motivates a distributed-barrier description. At low temperature, a fraction of the cluster moments is trapped in metastable configurations imposed by local anisotropy, disorder, and interactions with neighboring clusters. These moments contribute incompletely to the field-aligned magnetization within the finite field range and timescale of the hysteresis measurement. As temperature increases, a progressively larger fraction of the barrier distribution becomes dynamically accessible, allowing additional pre-existing moments to reach field-favored configurations during the measurement. Thus, the quantity that increases with temperature is the field-accessible contribution of the cluster ensemble, not the number of microscopic moments or the physical connectivity of the islands. This interpretation is consistent with the time-window dependence of blocked-particle magnetism and the evolution of hysteresis when thermally activated reversal becomes accessible.[55-59, 66]

For the more continuous film, the oxide forms a more strongly connected magnetic network. Most moments can contribute to the macroscopic response already at low temperature, so the dynamically restricted contribution is small. The temperature dependence is therefore dominated by thermal disorder and thermally assisted reversal, producing the conventional reductions in $M_s$ and $H_c$. The two samples can thus be treated as different limits of the same competition between intrinsic thermal demagnetization and dynamic access to a distributed set of reversal barriers.

In the phenomenological magnetization model, we represent the measured saturation magnetization as, $M_s(T) = M_0\left(1 - cT^{3/2}\right)[1 + \alpha\, g(T)]$, where the first factor describes the decrease of the intrinsic magnetic contribution with increasing temperature. The $T^{3/2}$ term is used only as a compact phenomenological representation of low-to-intermediate-temperature spin-wave-like demagnetization. The fitted coefficient $c$ should therefore be interpreted as an effective thermal-demagnetization parameter rather than as a direct measurement of bulk spin-wave stiffness. $M_0$ is the low-temperature magnetic scale for a given sample and incorporates the amount of oxide, the charge-transfer enhancement identified by DFT, and the population of magnetic defects or clusters.

The second factor in the $M_s$ equation accounts for the additional field-accessible contribution of moments whose response is dynamically restricted at low temperature. Here, $g(T)$ is the dynamically accessible fraction of these moments, and $\alpha$ is the relative magnitude of the magnetic-moment population whose field-aligned contribution is restricted at low temperature. A small α corresponds to a response that is already largely accessible at low temperature; whereas a large α signifies a substantial cluster contribution that becomes accessible across the measured temperature range.

Within a distributed-barrier picture, the dynamically accessible fraction *g(T)* can be written conceptually as:

$$g(T) = \int_0^{E_c(T)} P(E)dE\,, \qquad with\ E_c(T) = k_B T ln(\tau_m/\tau_0)$$

where *P(E)* is the distribution of effective reversal barriers, $\tau_m$ is the characteristic experimental observation time, and $\tau_0$ is the attempt time. The threshold $E_c(T)$ follows from the Néel-Brown condition $\tau(E,T) \leq \tau_m$.[55, 56] Because the present static magnetometry data do not independently determine *P(E)*, $\tau_m$, and $\tau_0$, we represent the cumulative response over the measured interval by the minimal normalized form: $g(t) = (T/T_{max})^n$. Here, the exponent *n* is an effective shape parameter for the cumulative barrier distribution sampled by the experiment, and controls how rapidly dynamic accessibility develops with temperature; $T_{max}$ is fixed at the upper experimental limit 350 K. Fixing $T_{max}$ removes an unconstrained activation-temperature parameter and does not introduce an additional fitted energy scale.

The empirical power law used for *g(T)* is not proposed as a universal microscopic barrier distribution. Rather, it is a minimal-parameter approximation to the monotonic cumulative accessibility expected for a broad barrier ensemble over a finite temperature window. Distributed-barrier and interaction-modified relaxation approaches are widely used in analyses of nanoparticle assemblies.[57-65]

On the other hand, coercivity $H_c$ is intrinsically sensitive to thermally activated reversal because a hysteresis loop records the field at which metastable magnetic states lose stability within a finite observation time. Thermal activation lowers the switching field relative to its zero-temperature value, as described by Néel-Brown and Sharrock-type treatments and by subsequent finite-temperature micromagnetic analyses.[55, 56, 66-68] For the continuous film, the observed monotonic softening is represented by $H_c(T) = H_{c0}\, e^{-(T/T_S)^r}$.

Here $H_{c0}$ is the extrapolated low-temperature coercive scale, $T_s$ is an effective thermal-softening scale—not a transition temperature—and $r$ controls the curvature of the softening. The stretched exponential is used phenomenologically to represent a distribution of reversal processes rather than a single barrier.

For the clustered film, the same distribution of anisotropy barriers that limits magnetic alignment also determines which clusters can switch their magnetization during the hysteresis cycle. In other words, the barrier distribution determines both the field-aligned magnetic response (how many cluster moments can access field-favored configurations) and the dynamic magnetic response (how rapidly and at what field the clusters can switch). The coercive field for the clustered film is therefore written as,

$$H_c(T) = [H_{base} + H_1\, g(T)]\, e^{-(T/T_s)^r}$$

The bracketed term describes the progressive involvement of dynamically accessible cluster moments in hysteretic reversal. $H_{base}$ is the residual low-temperature coercive contribution and $H_1$ measures the additional coercive scale associated with the cluster population entering the experimental time window. The exponential factor describes the competing thermal reduction of effective reversal barriers. At low and intermediate temperatures, the growth of $g(T)$ can dominate and increase $H_c$; at higher temperature, thermal softening dominates, producing a maximum followed by a decrease. Nonmonotonic coercivity can arise in size-distributed or interacting nanoparticle ensembles when different thermally activated reversal contributions compete, and numerical treatments of distributed assemblies reproduce such behavior.[64, 69]

In Figure 7c,d, the $M_s$ data for both samples were fitted simultaneously using a common thermal-demagnetization coefficient $c$ and a common exponent $n$, while $M_0$ and $\alpha$ were allowed to vary with morphology. The joint fit gives $c = (2.69 \pm 0.66) \times 10^{-5}$ $K^{-3/2}$ and $n = 0.569 \pm 0.269$. For the 1.5 nm sample, $M_0 = (45.49 \pm 0.88) \times 10^{-5}$ emu cm$^{-2}$ and $\alpha = 0.074 \pm 0.059$, with $R^2 = 0.923$. The small $\alpha$ indicates that nearly the entire measurable moment is accessible at low temperature; the fitted response is consequently governed mainly by conventional thermal demagnetization and reproduces the gradual decrease in $M_s$ above approximately 100 K.

For the 0.5 nm clustered film, $M_0 = (7.26 \pm 1.76) \times 10^{-5}$ emu cm$^{-2}$ and $\alpha = 1.23 \pm 0.47$, with $R^2 = 0.921$. The much larger $\alpha$ is the central result of the fit: the dynamically restricted contribution is comparable to the low-temperature magnetic scale, consistent with the broad distribution of barriers expected for the island morphology. The increase in *g(T)* overcompensates the intrinsic thermal-demagnetization term through most of the measured interval, yielding the observed increase in $M_s$. The common $n$ value provides a compact phenomenological description in which the two morphologies differ primarily in the magnitude of the dynamically restricted population rather than requiring unrelated temperature functions.

The continuous-film coercivity is reproduced by $H_{c0}$ = 174.7 ± 8.3 Oe, $T_s$ = 2355 ± 535 K, and $r$ = 0.431 ± 0.092, with $R^2$ = 0.990. The large effective $Ts$ and small $r$ produce broad, gradual softening, consistent with a connected magnetic film containing a distribution of reversal barriers. Importantly, $T_s$ is an empirical scale parameter of a stretched-exponential description rather than a physical transition temperature. Its large value reflects the gradual coercive softening observed over the experimental temperature interval and should not be interpreted as a transition or blocking temperature. The same data can be described equivalently over the measured range by a conventional sublinear power-law dependence of the form $H_c = H_0 - A \cdot T^p$.

The clustered-film fit gives $H_{base} \approx 0$, $H_1$ = 199.3 ± 31.1 Oe, $Ts$ = 374.7 ± 16.6 K, and $r$ = 3.94 ± 1.49, with $R^2$ = 0.963. Here, the $H_1$ *g(T)* term initially raises $H_c$ as more cluster moments become involved in thermally assisted switching and contribute to the hysteresis loop. At higher temperature, the rapidly increasing thermal-softening term becomes dominant and produces the downturn observed at 350 K. The model therefore reproduces both the rise and subsequent decrease using two physically motivated effects acting on the same barrier ensemble.

Taken together, the fitted parameters support a consistent morphology-dependent picture. Charge transfer from $MoSe_2$ enhances the local magnetic moments, while oxide morphology determines whether those moments form an already accessible connected response or a clustered ensemble with a broad distribution of effective barriers. The more continuous film is dominated by conventional thermal demagnetization and coercive softening. In the island film, thermally assisted access to metastable cluster configurations initially increases both the field-aligned magnetization and the population participating in magnetization switching; at the highest temperatures, thermal softening reduces $H_c$. This description links the AFM morphology, DFT charge-transfer result, and temperature-dependent hysteresis without assigning the anomalous temperature dependence to charge transfer.

## Conclusions

We demonstrate that ultrathin $VO_x/MoSe_2$ heterostructures exhibit enhanced magnetism governed by two complementary effects: interfacial electronic reconstruction and oxide morphology. Structural and spectroscopic measurements show a progression from discontinuous $VO_x$ islands toward more continuous films while preserving the underlying $MoSe_2$ lattice. XPS indicates negligible substitutional V incorporation into $MoSe_2$ and reveals a mixed-valence, nonstoichiometric $VO_x$ overlayer. First-principles calculations show that finite $VO_x$ clusters already possess magnetic moments and that interaction with $MoSe_2$ further enhances those moments through interfacial charge redistribution.

The thinnest films display an unusual increase in magnetization and a nonmonotonic coercivity with temperature, in contrast to the conventional thermal demagnetization and coercive softening of the more continuous film. A distributed-barrier phenomenological model captures this crossover by treating morphology as the factor that controls the collective accessibility and reversal dynamics of the pre-existing, charge-transfer-enhanced moments. The model therefore separates the roles of interface and morphology: electronic reconstruction enhances the local magnetic moment, whereas morphology governs its temperature-dependent collective response.

These results establish $VO_x/MoSe_2$ as an oxide/TMD platform in which magnetic functionality can be tuned without relying on substitutional doping of the semiconductor. More broadly, they show how interfacial charge transfer and nanoscale morphology can provide complementary control parameters for magnetic behavior in atomically thin heterostructures, with potential relevance to spintronic, magneto-optical, and multifunctional devices.

## Experimental Section

*CVD Synthesis of $MoSe_2$:* Si substrates with a 285 nm $SiO_2$ layer were cleaned sequentially with acetone, isopropanol, and deionized water before growth. Growth was performed in a 1 in. diameter quartz-tube reactor. As shown in Figure 1a, the $SiO_2$/Si substrates were positioned downstream in a 700-800 °C zone, while $MoSe_2$ powder (99.9%, Sigma-Aldrich) in a ceramic boat was placed upstream at the furnace center. The substrates were located 6-10 cm from the solid source. The furnace was heated to approximately 1040 °C over 60 min under $N_2$ flow (200 sccm), with the source and substrates initially outside the hot zone. When the furnace reached approximately 1035 °C, the source and substrates were moved to their growth positions by sliding the quartz tube. Simultaneously, water vapor was introduced by diverting the $N_2$ carrier gas through a bubbler containing 2 mL of deionized water at 24 °C. The $N_2/H_2O$ carrier flow was maintained for 8 min. After growth, the gas was switched to Ar containing 5% $H_2$ (200 sccm) for 4 min as a termination step.

*High-Vacuum Thermal Evaporation of $VO_x$:* Ultrathin $VO_x$ films were thermally evaporated onto CVD-grown $MoSe_2$ using an Angstrom Engineering high-vacuum evaporator equipped with a cryogenic pumping system. $V_2O_5$ was used as the evaporation source. The films were deposited at approximately 0.04 Å $s^{-1}$, and the nominal thickness was monitored with an INFICON quartz crystal sensor. Before deposition, the chamber was evacuated to approximately $10^{-7}$ Torr; during deposition, the pressure increased to approximately $10^{-6}$ Torr. The samples were maintained at room temperature.

*Raman and PL Spectroscopies*: Raman and photoluminescence measurements were performed with a confocal Raman microscope (LabRAM HR Evolution, Horiba Scientific) in backscattering geometry using 532 nm laser excitation. Raman and PL maps were acquired with a 50× objective while the sample remained stationary and the laser was scanned using the DuoScan™ imaging mode.

*XPS Measurements:* XPS spectra of as-prepared $MoSe_2$ and $MoSe_2$ coated with an ultrathin $VO_x$ layer were acquired in an ultrahigh-vacuum chamber equipped with a nonmonochromated dual-anode X-ray source and a Scienta R3000 hemispherical energy analyzer. The spectra reported here were collected using Al Kα radiation. Binding energies were calibrated to the $SiO_2$ Si 2p peak at 103.6 eV.

*Magnetic Measurements – Vibrating Sample Magnetometry*: Field- and temperature-dependent magnetic measurements were performed using the VSM option of a Quantum Design Physical Property Measurement System (PPMS). M-H loops were measured with the magnetic field applied both parallel and perpendicular to the sample plane, as indicated for the corresponding data sets, over the 10-350 K range discussed in this work. Temperature-dependent M–H loops were acquired sequentially upon cooling from 350 to 10 K. After each hysteresis measurement, the magnetic field was returned to zero and the sample was cooled in zero applied field to the next measurement temperature before acquiring the subsequent M–H loop. Saturation magnetization ($M_s$) was determined from the high-field plateau of the background-subtracted M–H loops, while the coercive field ($H_c$) was obtained from the zero-magnetization crossing of the hysteresis loop. Magnetization is reported per unit area. To estimate the actual $MoSe_2$-covered area for the discontinuous CVD films, large-area optical images were acquired using the collage-scanning mode of the LabRAM HR Evolution and analyzed with ImageJ; the resulting surface-coverage fraction was used in the areal normalization. Control measurements performed on clean $SiO_2$ substrates and pristine $MoSe_2/SiO_2$ samples using the same mounting and measurement procedures showed no detectable hysteresis within the experimental sensitivity. Together with the systematic $VO_x$-thickness and temperature dependence observed only in the heterostructures, these controls argue against a substrate-, holder-, or handling-related ferromagnetic contribution.

*Density Functional Theory Section:* We performed first-principles calculations using density functional theory with a Hubbard correction (DFT+U).[70] Calculations were carried out with the SIESTA code[71, 72] using the Perdew-Burke-Ernzerhof (PBE) generalized-gradient approximation (GGA) functional[73] and a double-ζ polarized (DZP) basis set. A mesh cutoff of 250 Ry, a Γ-centered 1 × 1 × 1 Monkhorst-Pack k-point grid[74] and a Hubbard parameter U = 4.0 eV for V were used.[75] A hexagonal 5 × 5 × 1 $MoSe_2$ supercell containing 75 atoms was first constructed using a previously optimized structure, with a = b = 16.75 Å, c = 30.00 Å, α = β = 90°, and γ = 120°. The optimized parameters were consistent with literature values,[76-79] supporting the computational setup. Two finite vanadium oxide clusters, $V_4O_{10}$ and $V_{10}O_{22}$, were then constructed from the bulk $V_2O_5$ structure and placed on the $MoSe_2$ supercell to examine their interfacial interaction. The $V_4O_{10}$ cluster contains the 14 atoms of one bulk $V_2O_5$ unit cell but is treated as a finite, nonperiodic cluster, with approximate dimensions of 2.54 × 9.22 × 5.34 $Å^3$. The $V_{10}O_{22}$ cluster is a 32-atom slab-like structure derived from bulk $V_2O_5$ cleaved along the (001) direction, with approximate dimensions of 8.67 × 3.30 × 10.18 $Å^3$. Hybrid structures were constructed by placing each cluster on the $MoSe_2$ supercell. These finite-cluster models are intended to represent local $VO_x/MoSe_2$ interactions relevant to the early stages of ultrathin-film growth rather than the full morphology of the experimental films.

**Acknowledgements**

H.R.G. acknowledges support from the National Science Foundation Grant No. NSF-ECCS-2430871, and from Nanobase Inc. Award: "Optical Properties of New Two-Dimensional Materials". H.R.G. thanks Prof. Matthias Batzill for kindly granting access to his laboratory facilities. B.I. and R.B.O. thank CNPq (process numbers 194155/2025-0 and 151043/2024-8, respectively). B.I. and G.S.L.F. thank FAPESP (process numbers 2024/11016-0 and 2024/03413-9, respectively), D.S.G. acknowledges the Center for Computing in Engineering and Sciences at UNICAMP for financial support through the FAPESP/CEPID grant 2013/08293-7. The authors also thank the Coaraci Supercomputer for computer time (FAPESP grant 2019/17874-0) and the Center for Computing in Engineering and Sciences at UNICAMP (FAPESP grant 2013/08293-4). D. S. G. also acknowledges support from INEO/CNPq and FAPESP grant 2025/27044-5.

**Conflict of interest**

The authors declare no conflict of interest.

**Author Contributions**

Y.W. and N.K. conducted synthesis experiments as well as Raman and PL spectroscopies with F.A.N. assistance. K.L. conducted XPS characterization. Y. W. performed VSM and AFM measurements, processed and analyzed the data with H.R.G. input. J.P. provided complementary AFM data. X.J. contributed with discussions and resources (high-vacuum thermal evaporation). R.B.O., B.I., G.S.L.F. and D.S.G. performed DFT calculations. H.R.G., Y.W. and N.K. conceived the project and wrote the manuscript with contributions from the other authors. All the authors read, commented on the manuscript, and gave their approval for the final manuscript.

**References**

(1) Burch, K. S.; Mandrus, D.; Park, J. G., Magnetism in two-dimensional van der Waals materials. *Nature* **2018,** *563* (7729), 47-52.

(2) Gong, C.; Li, L.; Li, Z. L.; Ji, H. W.; Stern, A.; Xia, Y.; Cao, T.; Bao, W.; Wang, C. Z.; Wang, Y. A.; Qiu, Z. Q.; Cava, R. J.; Louie, S. G.; Xia, J.; Zhang, X., Discovery of intrinsic ferromagnetism in two-dimensional van der Waals crystals. *Nature* **2017,** *546* (7657), 265.

(3) Huang, B.; Clark, G.; Navarro-Moratalla, E.; Klein, D. R.; Cheng, R.; Seyler, K. L.; Zhong, D.; Schmidgall, E.; McGuire, M. A.; Cobden, D. H.; Yao, W.; Xiao, D.; Jarillo-Herrero, P.; Xu, X. D., Layer-dependent ferromagnetism in a van der Waals crystal down to the monolayer limit. *Nature* **2017,** *546* (7657), 270.

(4) Wang, Q. H.; Kalantar-Zadeh, K.; Kis, A.; Coleman, J. N.; Strano, M. S., Electronics and optoelectronics of two-dimensional transition metal dichalcogenides. *Nat Nanotechnol* **2012,** *7* (11), 699-712.

(5) Xu, X. D.; Yao, W.; Xiao, D.; Heinz, T. F., Spin and pseudospins in layered transition metal dichalcogenides. *Nat Phys* **2014,** *10* (5), 343-350.

(6) Mak, K. F.; Lee, C.; Hone, J.; Shan, J.; Heinz, T. F., Atomically Thin $MoS_2$: A New Direct-Gap Semiconductor. *Phys Rev Lett* **2010,** *105* (13), 136805.

(7) Gutiérrez, H. R.; Perea-López, N.; Elías, A. L.; Berkdemir, A.; Wang, B.; Lv, R.; López-Urías, F.; Crespi, V. H.; Terrones, H.; Terrones, M., Extraordinary Room-Temperature Photoluminescence in Triangular $WS_2$ Monolayers. *Nano Lett* **2013,** *13* (8), 3447-3454.

(8) Sierra, J. F.; Fabian, J.; Kawakami, R. K.; Roche, S.; Valenzuela, S. O., Van der Waals heterostructures for spintronics and opto-spintronics. *Nat Nanotechnol* **2021,** *16* (8), 856-868.

(9) Mak, K. F.; Shan, J.; Ralph, D. C., Probing and controlling magnetic states in 2D layered magnetic materials. *Nat Rev Phys* **2019,** *1* (11), 646-661.

(10) Deng, J. J.; Zhou, Z. H.; Chen, J. L.; Cheng, Z. H.; Liu, J.; Wang, Z. Y., Vanadium-Doped Molybdenum Diselenide Atomic Layers with Room-Temperature Ferromagnetism. *Chemphyschem* **2022,** *23* (16), e202200162.

(11) Nisi, K.; Kiemle, J.; Powalla, L.; Scavuzzo, A.; Nguyen, T. D.; Taniguchi, T.; Watanabe, K.; Duong, D. L.; Burghard, M.; Holleitner, A. W.; Kastl, C., Defect-Engineered Magnetic Field Dependent Optoelectronics of Vanadium Doped Tungsten Diselenide Monolayers. *Adv Opt Mater* **2022,** *10* (17), 2102711.

(12) Pham, Y. T. H.; Liu, M. Z.; Jimenez, V. O.; Yu, Z. H.; Kalappattil, V.; Zhang, F.; Wang, K.; Williams, T.; Terrones, M.; Phan, M. H., Tunable Ferromagnetism and Thermally Induced Spin Flip in Vanadium-Doped Tungsten Diselenide Monolayers at Room Temperature. *Adv Mater* **2020,** *32* (45), 2003607.

(13) Yun, S. J.; Cho, B. W.; Dinesh, T.; Yang, D. H.; Kim, Y. I.; Jin, J. W.; Yang, S. H.; Nguyen, T. D.; Kim, Y. M.; Kim, K. K.; Duong, D. L.; Kim, S. G.; Lee, Y. H., Escalating Ferromagnetic Order via Se-Vacancies Near Vanadium in $WSe_2$ Monolayers. *Adv Mater* **2022,** *34* (10), 2106551.

(14) Yun, S. J.; Duong, D. L.; Ha, D. M.; Singh, K.; Phan, T. L.; Choi, W.; Kim, Y. M.; Lee, Y. H., Ferromagnetic Order at Room Temperature in Monolayer $WSe_2$ Semiconductor via Vanadium Dopant. *Adv Sci* **2020,** *7* (9), 1903076.

(15) Zhang, F.; Zheng, B. Y.; Sebastian, A.; Olson, D. H.; Liu, M. Z.; Fujisawa, K.; Pham, Y. T. H.; Jimenez, V. O.; Kalappattil, V.; Miao, L. X.; Zhang, T. Y.; Pendurthi, R.; Lei, Y.; Elias, A. L.; Wang, Y. X.; Alem, N.; Hopkins, P. E.; Das, S.; Crespi, V. H.; Phan, M. H.; Terrones, M., Monolayer Vanadium-Doped Tungsten Disulfide: A Room-Temperature Dilute Magnetic Semiconductor. *Adv Sci* **2020,** *7* (24), 2001174.

(16) Duan, H. L.; Guo, P.; Wang, C.; Tan, H.; Hu, W.; Yan, W. S.; Ma, C.; Cai, L.; Song, L.; Zhang, W. H.; Sun, Z. H.; Wang, L. J.; Zhao, W. B.; Yin, Y. W.; Li, X. G.; Wei, S. Q., Beating the exclusion rule against the coexistence of robust luminescence and ferromagnetism in chalcogenide monolayers. *Nature Communications* **2019,** *10*, 1584.

(17) Fu, S. C.; Kang, K.; Shayan, K.; Yoshimura, A.; Dadras, S.; Wang, X. T.; Zhang, L. H.; Chen, S. W.; Liu, N.; Jindal, A.; Li, X. Z.; Pasupathy, A. N.; Vamivakas, A. N.; Meunier, V.; Strauf, S.; Yang, E. H., Enabling room temperature ferromagnetism in monolayer $MoS_2$ via in situ iron-doping. *Nature Communications* **2020,** *11* (1), 2034.

(18) Chakraborty, S.; Ramesh, M.; Niranjan, M. K., Defect-engineered magnetism in conventional and Janus 2D vdW semiconducting monolayers: Benchmark insights from first-principles investigation. *J Magn Magn Mater* **2026,** *638*.

(19) Ge, J.; Luo, T. C.; Lin, Z. Z.; Shi, J. P.; Liu, Y. Z.; Wang, P. Y.; Zhang, Y. F.; Duan, W. H.; Wang, J., Magnetic Moments Induced by Atomic Vacancies in Transition Metal Dichalcogenide Flakes. *Adv Mater* **2021,** *33* (4).

(20) Zhao, C.; Norden, T.; Zhang, P. Y.; Zhao, P. Q.; Cheng, Y. C.; Sun, F.; Parry, J. P.; Taheri, P.; Wang, J. Q.; Yang, Y. H.; Scrace, T.; Kang, K. F.; Yang, S.; Miao, G. X.; Sabirianov, R.; Kioseoglou, G.; Huang, W.; Petrou, A.; Zeng, H., Enhanced valley splitting in monolayer $WSe_2$ due to magnetic exchange field. *Nat Nanotechnol* **2017,** *12* (8), 757.

(21) Zhong, D.; Seyler, K. L.; Linpeng, X. Y.; Cheng, R.; Sivadas, N.; Huang, B.; Schmidgall, E.; Taniguchi, T.; Watanabe, K.; McGuire, M. A.; Yao, W.; Xiao, D.; Fu, K. M. C.; Xu, X. D., Van der Waals engineering of ferromagnetic semiconductor heterostructures for spin and valleytronics. *Sci Adv* **2017,** *3* (5), e1603113.

(22) Ramos, L.; Ojo, A. I.; Wadumesthri, Y.; Almuhanna, I.; Gutierrez, H. R.; Arena, D. A., Exploring Interfacial Effects in Transition Metal Dichalcogenide/Ferrimagnetic Alloy Heterostructures. *Appl Sci-Basel* **2026,** *16* (10), 4828.

(23) Deng, Y. J.; Yu, Y. J.; Song, Y. C.; Zhang, J. Z.; Wang, N. Z.; Sun, Z. Y.; Yi, Y. F.; Wu, Y. Z.; Wu, S. W.; Zhu, J. Y.; Wang, J.; Chen, X. H.; Zhang, Y. B., Gate-tunable room-temperature ferromagnetism in two-dimensional FeGeTe. *Nature* **2018,** *563* (7729), 94.

(24) Hou, Y. S.; Wu, R. Q., Magnetic Anisotropy in 2D van der Waals Magnetic Materials and their Heterostructures: Importance, Mechanisms, and Opportunities. *Adv Funct Mater* **2025,** *35* (51).

(25) Gibertini, M.; Koperski, M.; Morpurgo, A. F.; Novoselov, K. S., Magnetic 2D materials and heterostructures. *Nat Nanotechnol* **2019,** *14* (5), 408-419.

(26) Choi, E. M.; Sim, K. I.; Burch, K. S.; Lee, Y. H., Emergent Multifunctional Magnetic Proximity in van der Waals Layered Heterostructures. *Adv Sci* **2022,** *9* (21), 2200186.

(27) Li, Z. X.; Qiu, L. X.; An, Y. P.; Zhan, Q. F.; Gong, S. J., Interface engineering and electric control of magnetocrystalline anisotropy in two-dimensional magnets. *Phys Rev B* **2026,** *113* (2), L020406.

(28) Mao, L.; Wang, X.; Huang, M. M.; Zhang, A. Y.; Chen, H. J.; Chen, W. Z. H.; Yi, A. J.; Peng, D.; Li, L. D.; Han, J. B., Interface Engineering Strategies for Magnetic and Magneto-Optical Enhancement of Two-Dimensional FeGeTe. *Acs Omega* **2026,** *11* (9), 14890-14897.

(29) Fabris, G. S. L.; Ipaves, B.; Oliveira, R. B.; Gutierrez, H. R.; Pereira , M. J. r.; Galvao, D. S., Magnetism Induced by Azanide and Ammonia Adsorption in Defective Molybdenum Disulfide and Diselenide: A First-Principles Study. *Acs Omega* **2026,** *11* (4), 6191-6197.

(30) Hellman, F.; Hoffmann, A.; Tserkovnyak, Y.; Beach, G. S. D.; Fullerton, E. E.; Leighton, C.; MacDonald, A. H.; Ralph, D. C.; Arena, D. A.; Dürr, H. A.; Fischer, P.; Grollier, J.; Heremans, J. P.; Jungwirth, T.; Kimel, A. V.; Koopmans, B.; Krivorotov, I. N.; May, S. J.; Petford-Long, A. K.; Rondinelli, J. M.; Samarth, N.; Schuller, I. K.; Slavin, A. N.; Stiles, M. D.; Tchernyshyov, O.; Thiaville, A.; Zink, B. L., Interface-induced phenomena in magnetism. *Rev Mod Phys* **2017,** *89* (2), 025006.

(31) Imada, M.; Fujimori, A.; Tokura, Y., Metal-insulator transitions. *Rev Mod Phys* **1998,** *70* (4), 1039-1263.

(32) Streltsov, S. V.; Khomskii, D. I., Covalent bonds against magnetism in transition metal compounds. *P Natl Acad Sci USA* **2016,** *113* (38), 10491-10496.

(33) Dagotto, E., Complexity in strongly correlated electronic systems. *Science* **2005,** *309* (5732), 257-262.

(34) Hu, P.; Hu, P.; Vu, T. D.; Li, M.; Wang, S. C.; Ke, Y. J.; Zeng, X. T.; Mai, L. Q.; Long, Y., Vanadium Oxide: Phase Diagrams, Structures, Synthesis, and Applications. *Chem Rev* **2023,** *123* (8), 4353-4415.

(35) Correal, S.; Hernández-Gómez, D.; Esquivel, A. S.; Cardona-Rodríguez, A.; Reiber, A.; Hernandez, Y.; González-Hernández, R.; Ramírez, J. G., Tuning electronic and magnetic properties through disorder in $V_2O_5$ nanoparticles. *Sci Rep-Uk* **2023,** *13* (1), 6752.

(36) Jia, L. N.; Hou, P. Z.; Liu, T. N.; Lu, P. P.; Liu, L. H.; Zhang, J. M.; Sun, H. Y., Room-temperature ferromagnetism and magnetization switching effect of ordered porous $V_2O_5$ nanofilms. *J Alloy Compd* **2025,** *1037*, 182353.

(37) Cezar, A. B.; Graff, I. L.; Varalda, J.; Schreiner, W. H.; Mosca, D. H., Oxygen-vacancy-induced room-temperature magnetization in lamellar $V_2O_5$ thin films. *J Appl Phys* **2014,** *116* (16), 163904.

(38) Subhan, F.; Kim, K.; Kim, J.; Qin, G. Z.; Ali, L.; Jang, J., van der Waals engineering of the valley polarization in $WSe_2$ the Kagome $V_2O_3$ monolayer heterostructure through magnetic proximity effect. *J Mater Chem C* **2026,** *14* (32), 14783-14794.

(39) Kapuruge, N.; Alba, T.; Lasek, K.; Schulz, N.; Wadumesthri, Y.; Nugera, F. A.; Jimenez, V. O.; Hyde, R.; Pan, J.; Srikanth, H.; Phan, M. H.; Gutierrez, H. R. *Hybrid 0D/2D Metamaterials with Room-Temperature Ferromagnetism*, in: 2023 IEEE Nanotechnology Materials and Devices Conference (NMDC), Paestum (Salerno), Italy, 2023; pp 817-818, DOI: 10.1109/NMDC57951.2023.10343993.

(40) Sharma, R.; Dang, D. T. X.; Woods, L. M., Magnetism at the interface of $MoSe_2/V_2O_5$ heterostructures. *J Phys Chem Solids* **2026,** *211*, 113403.

(41) Bedanta, S.; Kleemann, W., Supermagnetism. *J Phys D Appl Phys* **2009,** *42* (1), 013001.

(42) Sahoo, P. K.; Memaran, S.; Xin, Y.; Balicas, L.; Gutierrez, H. R., One-pot growth of two-dimensional lateral heterostructures via sequential edge-epitaxy. *Nature* **2018,** *553* (7686), 63.

(43) Sahoo, P. K.; Memaran, S.; Nugera, F. A.; Xin, Y.; Diaz Marquez, T.; Lu, Z.; Zheng, W.; Zhigadlo, N. D.; Smirnov, D.; Balicas, L.; Gutierrez, H. R., Bilayer Lateral Heterostructures of Transition-Metal Dichalcogenides and Their Optoelectronic Response. *ACS Nano* **2019,** *13* (11), 12372-12384.

(44) Nugera, F. A.; Sahoo, P. K.; Xin, Y.; Ambardar, S.; Voronine, D. V.; Kim, U. J.; Han, Y.; Son, H.; Gutierrez, H. R., Bandgap Engineering in 2D Lateral Heterostructures of Transition Metal Dichalcogenides via Controlled Alloying. *Small* **2022,** *18* (12), 2106600.

(45) Godin, K.; Cupo, C.; Yang, E. H., Reduction in Step Height Variation and Correcting Contrast Inversion in Dynamic AFM of $WS_2$ Monolayers. *Sci Rep-Uk* **2017,** *7*, 17798.

(46) Santos, S.; Verdaguer, A., Imaging Water Thin Films in Ambient Conditions Using Atomic Force Microscopy. *Materials* **2016,** *9* (3), 182.

(47) Butt, H. J.; Cappella, B.; Kappl, M., Force measurements with the atomic force microscope: Technique, interpretation and applications. *Surf Sci Rep* **2005,** *59* (1-6), 1-152.

(48) Nemes-Incze, P.; Osváth, Z.; Kamarás, K.; Biró, L. P., Anomalies in thickness measurements of graphene and few layer graphite crystals by tapping mode atomic force microscopy. *Carbon* **2008,** *46* (11), 1435-1442.

(49) Nam, D.; Lee, J. U.; Cheong, H., Excitation energy dependent Raman spectrum of $MoSe_2$. *Sci Rep-Uk* **2015,** *5*, 17113.

(50) Island, J. O.; Kuc, A.; Diependaal, E. H.; Bratschitsch, R.; van der Zant, H. S. J.; Heine, T.; Castellanos-Gomez, A., Precise and reversible band gap tuning in single-layer $MoSe_2$ by uniaxial strain. *Nanoscale* **2016,** *8* (5), 2589-2593.

(51) Cheng, X. R.; Jiang, L. Y.; Li, Y. Y.; Zhang, H. J.; Hu, C. S.; Xie, S. Y.; Liu, M.; Qi, Z. M., Using strain to alter the energy bands of the monolayer $MoSe_2$: A systematic study covering both tensile and compressive states. *Appl Surf Sci* **2020,** *521*, 146398.

(52) Druppel, M.; Deilmann, T.; Kruger, P.; Rohlfing, M., Diversity of trion states and substrate effects in the optical properties of an $MoS_2$ monolayer. *Nature Communications* **2017,** *8*, 2117.

(53) Ross, J. S.; Wu, S.; Yu, H.; Ghimire, N. J.; Jones, A. M.; Aivazian, G.; Yan, J.; Mandrus, D. G.; Xiao, D.; Yao, W.; Xu, X., Electrical control of neutral and charged excitons in a monolayer semiconductor. *Nature Communications* **2013,** *4*, 1474.

(54) Han, H. V.; Lu, A. Y.; Lu, L. S.; Huang, J. K.; Li, H.; Hsu, C. L.; Lin, Y. C.; Chiu, M. H.; Suenaga, K.; Chu, C. W.; Kuo, H. C.; Chang, W. H.; Li, L. J.; Shi, Y., Photoluminescence Enhancement and Structure Repairing of Monolayer $MoSe_2$ by Hydrohalic Acid Treatment. *ACS Nano* **2016,** *10* (1), 1454-61.

(55) Néel, L., Théorie du traînage magnétique des ferromagnétiques en grains fins avec application aux terres cuites. *Annales de Géophysique* **1949,** *5*, 99–136.

(56) Brown, W. F., Thermal Fluctuations of a Single-Domain Particle. *Phys Rev* **1963,** *130* (5), 1677.

(57) Dormann, J. L.;  Fiorani, D.; Tronc, E., Magnetic relaxation in fine-particle systems. *Advances in Chemical Physics* **1997,** *98*, 283-494.

(58) Coffey, W. T.; Kalmykov, Y. P., Thermal fluctuations of magnetic nanoparticles: Fifty years after Brown. *J Appl Phys* **2012,** *112* (12), 121301.

(59) Stoner, E. C.; Wohlfarth, E. P., A Mechanism of Magnetic Hysteresis in Heterogeneous Alloys. *Philos Tr R Soc S-A* **1948,** *240* (826), 599-642.

(60) Papaefthymiou, G. C., Nanoparticle magnetism. *Nano Today* **2009,** *4* (5), 438-447.

(61) Dormann, J. L.;  Fiorani, D.; Tronc, E., On the models for interparticle interactions in nanoparticle assemblies: comparison with experimental results. *J Magn Magn Mater* **1999,** *202* (1), 251-267.

(62) Dormann, J. L.;  DOrazio, F.;  Lucari, F.;  Tronc, E.;  Prene, P.;  Jolivet, J. P.;  Fiorani, D.;  Cherkaoui, R.; Nogues, M., Thermal variation of the relaxation time of the magnetic moment of gamma-$Fe_2O_3$ nanoparticles with interparticle interactions of various strengths. *Phys Rev B* **1996,** *53* (21), 14291-14297.

(63) Dormann, J. L.;  Bessais, L.; Fiorani, D., A Dynamic Study of Small Interacting Particles - Superparamagnetic Model and Spin-Glass Laws. *J Phys C Solid State* **1988,** *21* (10), 2015-2034.

(64) Lee, J. S.;  Tan, R. P.;  Wu, J. H.; Kim, Y. K., Effect of interparticle interactions and size dispersion in magnetic nanoparticle assemblies: A static and dynamic study. *Appl Phys Lett* **2011,** *99* (6), 062506.

(65) Morup, S.; Tronc, E., Superparamagnetic Relaxation of Weakly Interacting Particles. *Phys Rev Lett* **1994,** *72* (20), 3278-3281.

(66) Sharrock, M. P., Time-Dependence of Switching Fields in Magnetic Recording Media (Invited). *J Appl Phys* **1994,** *76* (10), 6413-6418.

(67) Suess, D.;  Vogler, C.;  Abert, C.;  Bruckner, F.;  Windl, R.;  Breth, L.; Fidler, J., Fundamental limits in heat-assisted magnetic recording and methods to overcome it with exchange spring structures. *J Appl Phys* **2015,** *117* (16), 163913.

(68) Okamoto, S., Experimental approaches for micromagnetic coercivity analysis of advanced permanent magnet materials. *Sci Technol Adv Mat* **2021,** *22* (1), 124-134.

(69) Iglesias, O.; Labarta, A., Magnetic relaxation in terms of microscopic energy barriers in a model of dipolar interacting nanoparticles. *Phys Rev B* **2004,** *70* (14), 144401.

(70) Kohn, W.; Sham, L. J., Self-Consistent Equations Including Exchange and Correlation Effects. *Phys Rev* **1965,** *140* (4A), A1133.

(71) García, A.;  Papior, N.;  Akhtar, A.;  Artacho, E.;  Blum, V.;  Bosoni, E.;  Brandimarte, P.;  Brandbyge, M.;  Cerdá, J. I.;  Corsetti, F.;  Cuadrado, R.;  Dikan, V.;  Ferrer, J.;  Gale, J.;  García-Fernández, P.;  García-Suárez, V. M.;  García, S.;  Huhs, G.;  Illera, S.;  Korytár, R.;  Koval, P.;  Lebedeva, I.;  Lin, L.;  López-Tarifa, P.;  Mayo, S. G.;  Mohr, S.;  Ordejón, P.;  Postnikov, A.;  Pouillon, Y.;  Pruneda, M.;  Robles, R.;  Sánchez-

Portal, D.; Soler, J. M.; Ullah, R.; Yu, V. W. Z.; Junquera, J., Siesta: Recent developments and applications. *J Chem Phys* **2020,** *152* (20), 204108.

(72) Soler, J. M.; Artacho, E.; Gale, J. D.; García, A.; Junquera, J.; Ordejón, P.; Sánchez-Portal, D., The SIESTA method for order-N materials simulation. *J Phys-Condens Mat* **2002,** *14* (11), 2745-2779.

(73) Perdew, J. P.; Burke, K.; Ernzerhof, M., Generalized gradient approximation made simple. *Phys Rev Lett* **1996,** *77* (18), 3865-3868.

(74) Monkhorst, H. J.; Pack, J. D., Special Points for Brillouin-Zone Integrations. *Phys Rev B* **1976,** *13* (12), 5188-5192.

(75) Smirnov, M. B.; Roginskii, E. M.; Smirnov, K. S.; Baddour-Hadjean, R.; Pereira-Ramos, J. P., Unraveling the Structure-Raman Spectra Relationships in $V_2O_5$ Polymorphs via a Comprehensive Experimental and DFT Study. *Inorg Chem* **2018,** *57* (15), 9190-9204.

(76) Guguchia, Z.; Kerelsky, A.; Edelberg, D.; Banerjee, S.; von Rohr, F.; Scullion, D.; Augustin, M.; Scully, M.; Rhodes, D. A.; Shermadini, Z.; Luetkens, H.; Shengelaya, A.; Baines, C.; Morenzoni, E.; Amato, A.; Hone, J. C.; Khasanov, R.; Billinge, S. J. L.; Santos, E.; Pasupathy, A. N.; Uemura, Y. J., Magnetism in semiconducting molybdenum dichalcogenides. *Sci Adv* **2018,** *4* (12), eaat3672.

(77) Haldar, S.; Vovusha, H.; Yadav, M. K.; Eriksson, O.; Sanyal, B., Systematic study of structural, electronic, and optical properties of atomic-scale defects in the two-dimensional transition metal dichalcogenides $MX_2$( M= Mo, W; X= S, Se, Te). *Phys Rev B* **2015,** *92* (23), 235408.

(78) Devi, A.; Dhiman, N.; Kumar, N.; Alfalasi, W.; Kumar, A.; Ahluwalia, P. K.; Singh, A.; Tit, N., Ferromagnetism in Defected TMD ($MoX_2$, X = S, Se) Monolayer and Its Sustainability under $O_2$, $O_3$, and $H_2O$ Gas Exposure: DFT Study. *Nanomaterials-Basel* **2023,** *13* (10), 1642.

(79) Ma, Y. D.; Dai, Y.; Guo, M.; Niu, C. W.; Lu, J. B.; Huang, B. B., Electronic and magnetic properties of perfect, vacancy-doped, and nonmetal adsorbed $MoSe_2$, $MoTe_2$ and $WS_2$ monolayers. *Phys Chem Chem Phys* **2011,** *13* (34), 15546-15553.